\documentclass{article}

\usepackage{epsfig}

\begin{document}

\title{\hspace{-2.5cm}{\small Horizons in Superconductivity Research (ed. F. Columbus, Nova
Science, New York, 2003)}\\ \vskip 0cm {\bf On Critical
Current Enhancement in Dislocated, Deoxygenated and Irradiated
Superconductors:\\ A Unified Approach}} \vspace{-1.0cm}
\author{SERGEI SERGEENKOV}
\maketitle \vspace{-1.20cm}
\begin{center}
{\large Bogoliubov Laboratory of Theoretical
 Physics,\\ Joint Institute for Nuclear Research, 141980 Dubna,
 Russia}
\end{center}

\section{Introduction}
As is well-known~\cite{rev3}, the technological usage of any
superconducting materials is based on their ability to carry (without
loss) significant critical currents in strong enough applied magnetic
fields. This ability is directly related to pinning efficiency of a
given material which in turn is determined by its crystallographic
structure and the presence of different kinds of defects (both
inherent and introduced artificially). In conventional
superconductors, the magnetic flux flowing through the crystal is
assumed to be pinned by practically {\it immobile} (frozen) pinning
centers (except perhaps for a possibility of thermal fluctuations
around their equilibrium positions). In high-$T_c$ superconductors
(HTS) the situation is much more complicated because of the smallness
of their coherence length and, as a result, of practically inevitable
formation of intricate weak-link structure even within a single grain
(the so-called intragranular granularity~\cite{rev0}). Hence, any
defects (imperfections) in these materials will contribute not only
to their flux pinning ability but will also determine their weak-link
properties. Such dualism brings about a lot of interesting anomalies
in HTS (for the recent reviews on the subject, see,
e.g.,~\cite{rev1,rev2} and further references therein) and calls for
non-traditional pinning scenarios capable of explaining the observed
non-trivial electronic transport behavior in these materials. In the
present paper, one of the possible scenarios is proposed based on a
novel concept of vortex pinning via defect-induced intragrain weak
links which takes advantage of the above-mentioned dualism of
(extended) defects (as pinning centers and weak links) in HTS by
allowing pinning centers to participate in the pinning process more
actively. By considering a subtle balance between different forces
acting upon an extended (dislocations) and point (oxygen vacancies)
defects to stabilize their equilibrium position inside a crystal, the
scenario allows us to introduce external fields which control the
defect behavior of the superconducting sample and, as a result,
enables to describe (at least qualitatively) a rather wide set of the
observed anomalous properties of HTS crystals (see Section 2). The
paper is organized as follows. In Section 2 we review some of the key
experimental results which will be discussed within the proposed
scenario (outlined in Section 3) in the next sections. In Section 4
we consider pinning force improvement in screw dislocated thin films
(and related electric field induced phenomena). Section 5 deals with
the so-called fishtail anomaly (or peak effect) in oxygen-depleted
single crystals and stoichiometric melt-textured materials. In
Section 6 we demonstrate a substantial critical current density
enhancement in particle irradiated crystals. And finally, in Section
7 we summarize the obtained results and discuss the important
consequences of the proposed here unified scenario for optimization
of some application oriented vital characteristics of high-$T_c$
superconductors.

\section{Review of the key experimental results}

Recall~\cite{1} that there are serious arguments to consider the
twinning boundary (TB) in HTS as insulating regions of the Josephson
SIS-type structure. In particular, intragranular  weak  links due to
the structural defects have been discussed by  Schnelle  et
al.~\cite{2} and Hervieu et al.~\cite{3}. They observed antiphase
boundaries, twinning, and "tilt" boundaries, and considered
structural models for these different boundaries  as  well  as  their
influences  upon HTS properties. Additional evidence in favor of this
conclusion is due to observed deviations in stoichiometry at the TB.
As was shown by Zhu et al.~\cite{4} for
$YBa_2(Cu_{1-x}M_x)_3O_{7-\delta }$ samples, the TB thickness is
changed from $7\dot A$ (for $M=Ni,x=0.02,$ and $\delta =0$) to
$26\dot A$ (for $M=Al,x=0.02,$ and $\delta =0$), while for a pure
$YBa_2Cu_3O_7$ it gives $\simeq 10\dot A$. If after Zhu et
al.~\cite{5} we assume that the physical thickness of the TB is of
the order of an interplane distance (atomic scale), then such a
boundary should rather rapidly move via the movement of the twinning
dislocations. And this type of motion indeed has been observed by the
electron-microscopic images in HTS~\cite{6,7,8}. As a result of the
structural phase transition (if special precautions have not been
done) a HTS sample is divided into a large number of twinning
blocks~\cite{9}, i.e. the sample is crossed by a large number of
insulating layers (twin boundaries). An average distance between
boundaries is essentially less than the grain size. This net of
layers induces the net of Josephson junctions inside a single grain.
In addition, the well known processes of the oxygen ordering in HTS
leads to the continuous change of the lattice period along TB with
the change of the oxygen content~\cite{10,11}. Babcock and
Larbalestier~\cite{13} observed the regular networks of localized
grain boundary dislocations (GBDs) with the spacing ranged from $10
nm$ to $100 nm$. They concluded that closely spaced GBDs may produce
effectively continuous normal or insulating barriers at the boundary
which then behave like a planar SNS or SIS junction.  Cai et
al.~\cite{a6} observed pronounced peaks in the critical-current
density in $YBa_2Cu_3O_7$ bicrystals well above the lower critical
field. These peaks correspond to the fields for which the spacing
between intragrain vortices is commensurate with the wavelength of
the periodic grain boundary facet structure observed in the same
bicrystals. In turn, Tsu et al.~\cite{b9} made an important
comparison of grain boundary topography and dislocation network
structure in bulk-scale [001] tilt bicrystals of $Bi_2Sr_2CaCu_2O_8$
and $YBa_2Cu_3O_7$. At the same time, Diaz et al.~\cite{b2} and Shang
et al.~\cite{14} found clear evidence for vortex pinning by
dislocations in $YBa_2Cu_3O_7$ low-angle grain boundaries and
observed dislocations inside the grains along with coherent
intragranular boundaries acting as weak links. By imaging epitaxial
$YBa_2Cu_3O_{7-\delta }$ films with scanning tunneling microscopy,
Gerber et al.~\cite{15} have directly observed screw dislocations
with densities of $\simeq 10^9cm^{-2}$ sufficient to account for a
major part of the pinning forces and for weak links contribution in
their films. Chen et al.~\cite{16} found the strong correlation
between high density of the TBs and critical current density
enhancement in $YBa_2Cu_3O_{7-\delta }$ films. A more direct evidence
as for influence of dislocations on the pinning ability of HTS has
been obtained by Mannhart et al.~\cite{17} (and recently confirmed by
Hlubina et al.~\cite{b16}) in screw dislocated HTS single crystals.
They detected a nearly linear increase of the critical current
density with the number of dislocations as well as a substantial
improvement of $j_c(H)$ in applied magnetic field (at least, up to
$0.05 T$). The use of the same screw dislocated thin films in the
so-called superconducting field-effect transistor (SuFET) devices
allowed Mannhart et al.~\cite{18} to detect the electric-field
induced modulation of critical current density in their films.
Depending on the gate polarity, a quite tangible change (increase or
decrease) of $j_c(E)$ has been observed.

Recent STM-based imaging of the granular structure in underdoped
crystals of $Bi_2Sr_2CaCu_2O_{8+\delta}$~\cite{20a} revealed an
apparent segregation of its electronic structure into superconducting
domains (of the order of a few nanometers) located in an
electronically distinct background. In particular, it was found that
at low levels of hole doping ($\delta
<0.2$), the holes become concentrated at certain hole-rich domains.
Tunneling between such domains leads to intrinsic granular
superconductivity (GS) in HTS. Probably one of the first examples of
GS was observed in $YBa_2Cu_3O_{7-\delta }$ single crystals in the
form of the so-called fishtail anomaly of
magnetization~\cite{19,20,21,22}. In particular, Yang et
al.~\cite{19} found out that dislocation networks provide effective
pinning centers when the characteristic length scale of network
matches that of the vortex spacing. They stressed that in high-$T_c$
superconductors, oxygen vacancies are effective flux-pinning centers
at low temperatures while the high temperature spatial variation of
the oxygen ordering may well account for the anomalous fishtail
magnetization feature in $YBa_2Cu_3O_{7-\delta }$ crystals. Daeumling
et al.~\cite{20} discussed a rather intriguing correlation between
defect pinning, intragrain weak links, and oxygen deficiency in
$YBa_2Cu_3O_{7-\delta }$ single crystals. In particular, they have
found that as the nominal oxygen deficiency $\delta $ decreases
towards zero, the flux pinning declines and the crystals lose their
explicitly granular signature. The above anomalous magnetic-field
behavior has been attributed to the field-induced intragrain
granularity in oxygen-depleted materials~\cite{21}. A phase diagram
$H_m(\delta ,T)$ that demarcates the multigrain onset as a function
of temperature and oxygen deficiency reconstructed by Osofsky et
al.~\cite{22} allowed them to confirm that their single crystals
exhibited behavior characteristic of homogeneous superconductors for
$H<H_m$ and inhomogeneous superconductors for $H>H_m$. The granular
behavior for $H>H_m$ has been related to the clusters of oxygen
defects (within the $CuO$ plane) that restrict supercurrent flow and
allow excess flux to enter the crystal. The field at which the
fishtail anomaly occurs was found~\cite{23,24,23a} to decrease both
with increasing the temperature and oxygen deficiency. This in turn
suggests that the characteristic scale of the defect network
structure is strongly dependent on oxygen stoichiometry. On the other
hand, Ullrich et al.~\cite{25,26} have observed a rather substantial
critical current enhancement in stoichiometric melt-textured
crystallites with the dislocation density of $\cong $ $2\times
10^{10}cm^{-2}$. They argue that the strained region surrounding a
dislocation can cause the fishtail anomaly as well. Indeed, since the
strained regions near the dislocation core as well as
oxygen-deficient regions are both result in a lower $B_{c2}$ (the
upper critical field) compared to the $YBa_2Cu_3O_{7-\delta }$
matrix, they become (normal conducting) pinning centers if the
external magnetic field exceeds their upper critical field. The
microstructure of subgrain boundaries occurring in single-domain
$RBa_2Cu_3O_7$ ($R = Y$ and $Nd$) melt-textured composites has been
studied by Sandiumenge et al.~\cite{a7} using transmission electron
microscopy. It was found that subgrain boundaries (SGB's) have a
strong tendency to develop parallel to the (100), (010), and (110)
planes, while the form of dislocation networks is controlled by the
properties of constituting dislocations. Besides the underlying
dislocation networks, SGB's may develop mesostructures such as
faceting and stepped interfaces accommodating the deviation from
low-index planes. The scaling and relaxation behavior around the
fishtail minimum was studied in detail by Jirsa et al.~\cite{a2} in a
wide temperature range on $DyBa_2Cu_3O_7$ single crystals exhibiting
a pronounced fishtail effect.  It is also important to mention a
recent paper of Banerjee et al.~\cite{b15} who investigated evolution
of the peak effect with defect structure in $YBa_2Cu_3O_7$ thin films
at microwave frequencies.

Another important way to improve pinning ability of HTS single
crystals is to use irradiation to incorporate (in a controllable way)
different types of defects into these materials. Civale et
al.~\cite{27,28} achieved the pinning enhancement at high fields and
high temperature for vortex confinement by columnar defects in
$YBa_2Cu_3O_{7-\delta }$ crystals. Aligned columns of damaged
material $50\dot A$ in diameter and $15\mu m$ long were produced by
$580 Mev$ $Sn$ ion irradiation. The pinning obtained was much greater
than that produced by random point defects, and caused a considerable
enlargement of the irreversibility region in the $H-T$ plane. Hardy
et al.~\cite{29,30} have measured magnetic properties of the $Pb$ ion
irradiated $YBa_2Cu_3O_{7-\delta }$ crystals and found a rather
strong fluence dependence of irradiation-induced critical current
enhancement $j_c(\Phi ,H)$. The maximum of $j_c(\Phi ,H)$ was found
to shift to higher fluences with increasing the applied field and to
higher fields with increasing the irradiation fluence. An important
correlation between the extent of radiation damage and the oxygen
stoichiometry of the sample has been observed by Zhu et
al.~\cite{31}. They found that decrease of $\delta $ significantly
reduces the radiation damage. More evidence in favor of this
conclusion has been provided by Li et al.~\cite{b14}. They observed a
self-doping caused by oxygen displacements in heavy-ion-irradiated
$Bi_2Sr_2CaCu_2O_8$ crystals. Specially designed heavy-ion
irradiation experiments have been carried out for single crystals of
$YBa_2Cu_4O_8$ by Ming Xu et al.~\cite{a5} in order to compare the
magnetic-field dependence of the flux pinning for damage channels
with the intrinsic fishtail pinning that occurs in this material
before irradiation. The two pinning effects seem to be additive in
that irradiation increases the hysteresis at all fields and the
fishtail hump persists in the irradiated samples with roughly the
same magnitude as before the damage channels are produced. For
irradiated samples, there are two regions where pinning rises with
increasing field, one near zero field due to a matching effect of the
defect density with the vortex lattice spacing and a second at the
same field as the original fishtail. Both effects appear in the same
sample in different magnetic-field ranges. In order to assess the
dependence of the fishtail on the defect size and concentration,
Werner et al.~\cite{a1} modified the defect structure by reactor
neutron irradiation and additional annealing treatments. Their
results emphasize the important role of normal conducting regions,
which are created by clustering of defects, typically of oxygen
vacancies. Non-Ohmic resistive state in $120Mev$ $^{16}O$ ion-%
irradiated $YBa_{2}Cu_{3}O_{7-\delta }$ has been studied by Iwase et
al.~\cite{x1}. They found a strong influence of the irradiation dose
of fluence $\Phi $  on the power-like current-voltage characteristics
$V\propto I^{n}$ with $n=f(T)\log (\Phi /\Phi _{0})$. The effect of
$2.5Mev$  electron  irradiation  on ceramic sintered
$YBa_{2}Cu_{3}O_{7-\delta }$ was  investigated  by Gilchrist  and
Konczykowski~\cite{z1}. They deduced a decrease of  the
characteristic magnetic field of Josephson junctions after
irradiation and a very slight increase of intergrain critical current
at very low  doses. The authors noted that the accommodation of
defects in the barrier region can lead to junction degradation and
the loss of  Josephson tunneling capability. More recently, $He^+$
irradiation effects on $YBa_2Cu_3O_{7-\delta}$ GBJJs modified by
oxygen annealing have been studied by Navacerrada et
al.~\cite{b7,b8}. At the same time, Klaassen et al.~\cite{a8} argue
that natural linear defects in thin films form an analogous system to
columnar tracks in irradiated samples. There are, however, three
essential differences: (i) typical matching fields are at least one
order of magnitude smaller, (ii) linear defects are smaller than
columnar tracks, and (iii) the distribution of natural linear defects
is nonrandom, whereas columnar tracks are randomly distributed.

\section{Dislocation induced Josephson effect}

To realize a scenario for the dislocation-induced atomic scale
Josephson effect in HTS single crystals, let us consider
the model of small Josephson contacts~\cite{39} with length
$L<\lambda _J$ ($\lambda _J=%
\sqrt{\phi _0/\mu _0lj_{c0}}$ is the Josephson penetration depth), in
a strong enough magnetic field (which is applied normally to the
contact area $l\times L$) such that $H>\phi _0/2\pi \lambda _Jl$,
where $l=\lambda _{L1}+\lambda _{L2}+t$ ($\lambda _{L1(2)}$ is the
London penetration depth of the first (second) junction, and $t$ is
an insulator thickness). Assuming that twinning dislocations, lying
along the z-axis (which coincides with the direction of the applied
magnetic field $H$) and distributed along the x-axis with
dimensionless density $\rho (x)$, give rise to an (inhomogeneous)
Josephson supercurrent density $j_c[\rho (x)]$, the critical current
density through such an inhomogeneous single Josephson contact
reads~\cite{s1,40}
\begin{equation}
j_s^J(\rho ,H)=\frac 1L\int\limits_0^Ldxj_c[\rho (x)]\sin (kx+c).
\end{equation}
Here $k=2\pi lH/\phi _0$, $c$ is the zero-field phase difference
between two superconductors forming a SIS-type Josephson contact, and
we have assumed that the length of the contact $L$ is the twin
boundary (TB) length and the insulator thickness $t$ corresponds to
the TB thickness. Notice that the insulator thickness of the
dislocation-induced Josephson junction, $t(x)$, is related to $\rho
(x)$ as follows~\cite{40,41,42}
\begin{equation}
t(x)=\int\limits_x^Ldx^{\prime }\rho (x^{\prime })
\end{equation}
Applicability of the dislocation description of a thin TB is
known~\cite{9,41} to be restricted by the condition $t\ll L.$ As we
shall see below, this condition is reasonably satisfied in dislocated
HTS crystals. To proceed further, we postulate a $\delta$-functional
form of dislocation-induced critical current density assuming that
$j_c[\rho (x)]$ is centered around an average (dimensionless)
dislocation density $\rho $
\begin{equation}
j_c[\rho (x)]=j_{c0}e^{-t(x)/\xi _n}\delta [\rho (x)-\rho ]
\end{equation}
Here $j_{c0}$ obeys the usual Ambegoakar-Baratoff expression for
Josephson critical current~\cite{39}, the exponential dependence
accounts for the probability of Josephson tunneling on the local TB
of thickness $t(x)$, and $\xi _n=\hbar v_F/\pi k_BT$ is the decay
length of the contact. Taking into account the known property of the
Dirac delta function $\delta [\rho
(x)-\rho ]=\delta [x-x_0(\rho )]/{\mid {\rho ^{\prime }(x)}\mid }%
_{x=x_0(\rho )}$ where ${{\rho ^{\prime }(x)\equiv d\rho /dx,}}$ and $%
x_0(\rho )$ is the solution of the equation $\rho (x_0)=\rho $,
Eqs.(1)-(3) give for the dislocation-induced single-junction critical
current density
\begin{equation}
j_s^J(\rho ,H)=j_s(\rho ,0)\sin \left( \frac H{H_\rho ^J}+c\right),
\end{equation}
where 
\begin{equation}
j_s(\rho ,0)= j_{c0}\frac{\exp[-t(x_0)/\xi _n]}{L|\rho ^{\prime
}(x_0)|},
\end{equation}
and
\begin{equation}
 H_\rho ^J=\left (\frac{L}{x_0}\right )H_0^J.
\end{equation}
Here $x_0=x_0(\rho )$ and $H_0^J=\phi _0/2\pi lL$ is a characteristic
Josephson field.

To model a real situation of an inhomogeneous (due to the intragrain
granularity) single crystal, let us consider a random network of
dislocation-induced Josephson junctions (JJ). Assuming, for
simplicity, that
JJ are randomly distributed according to the exponential law $%
P(l)=(1/l_0)\exp (-l/l_0)$ with $l_0$ being an average "sandwich"
thickness, Eq.(4) results in the maximum (with $c=\pi /2$) critical
current density for a granular superconductor~\cite{43}:
\begin{equation}
j_s(\rho ,H)=\int\limits_0^\infty dlP(l)j_s^J(\rho ,H)=\frac{j_s(\rho ,0)}{%
1+H^2/H_\rho ^2},
\end{equation}
where $H_\rho =\phi _0/2\pi l_0x_0(\rho )$. As is seen, Eq.(7), at
least formally, describes the ordinary decrease of the critical
current density with $H$. But actually this is not the case in view
of the implicit dependence of a characteristic Josephson field
$H_\rho $ on the number of dislocations and on external sources
($\omega _0$, see below) affecting the distribution of defects inside
a crystal.

\section{Screw dislocated thin films}
\subsection{Critical current density versus dislocation density}

Supposing, for simplicity, the twinning boundaries (TB) as the only
weak links sources, we are able to use for their treatment the
dislocation model of elastic twinning~\cite{41} (see also the
comprehensive review~\cite{9} and references therein). According to
this model, the distribution of dislocations with density $\rho (x)$
along the elastic thin twin boundary is defined by the equilibrium
condition under the influence of the external forces $\omega (x)$
\begin{equation}
\int_0^L\frac{dx^{\prime }}{x^{\prime }-x}\rho (x^{\prime })=\omega
(x),
\end{equation}
where
\begin{equation}
\omega (x)=\frac{2\pi f(x)}{\mu b}.
\end{equation}
Here, $L$ is the twinning length, $\mu $ is the shear modulus of a
dislocation, $b$ is the magnitude of the Burgers vector, and $f(x)$
is a force affecting the distribution of dislocations within the TB.
To make our consideration more definitive, let us discuss a concrete
profile of the TB, i.e. solve the integral equation (8). If the
dislocations are constrained only by external forces (the so-called
freely growing TB), then the dislocation density distribution of such
a growing TB is governed by the law~\cite{44}
\begin{equation}
\rho (x)=-\frac 1{\pi ^2}\sqrt{x(L-x)}\int_0^L\frac{dx^{\prime
}}{x^{\prime }-x}\frac{\omega (x^{\prime })}{\sqrt{x^{\prime
}(L-x^{\prime })}}
\end{equation}
In particular, for the case of the homogeneous external forces (with
$\omega (x)\equiv \omega _0$), the above distribution near the TB tip
reads (an example of inhomogeneous external forces, resulting in a
self-tuned Josephson effect, is considered in \cite{40})
\begin{equation}
\rho (x)=\rho _0\sqrt{1-\frac xL},\qquad \rho _0\equiv \frac{\omega
_0}\pi
\end{equation}
The above expression is valid for $a_0\le x\le L$, where $a_0$ is of
the order of interatomic spacing. To describe correctly the vicinity
of the TB origin ($x<a_0$), a more accurate treatment of Eq.(8) is
needed. In view of Eqs.(5) and (6), distribution (11) results in the
following explicit $\rho$ dependencies of the zero-field critical
current density
\begin{equation}
j_s(\rho ,0)=2j_{c0}\left (\frac {\rho}{\rho _0^2}\right )\exp\left
[- \left (\frac{b}{\xi _n} \right )\left (\frac{\rho}{\rho _0}\right
)^3\right ] ,
\end{equation}
and characteristic Josephson field
\begin{equation}
H_\rho =H_0\frac{\rho _0}{1-(\rho /\rho _0)^2},
\end{equation}
respectively, where $H_0=\phi _0/3\pi bl_0$. So, in view of Eq.(12),
$j_s(\rho ,0)$ increases with $\rho$ for $\rho
<\rho _m$, has a maximum at $\rho =\rho _m\equiv \rho _0\sqrt[3]{\xi
_n/3b}$, and then exponentially decreases for $\rho
>\rho _m$. Using HTS parameters, $v_F\simeq 2\times 10^5m/s$, $T_c\simeq
90K$, and $b\simeq 1nm$, we get $\rho _m\simeq 1.2\rho _0$. On the
other hand, according to the validity of distribution (11) and taking
into account that usually $a_0/L<10^{-2}$, we find that Eqs.(12) and
(13) are valid for $0\le \rho /\rho _0<1$ only. It means that up to
the highest admissible dislocation density (i.e., up to $\rho \simeq
\rho _0$) we are always in the linear region (with $\rho <\rho _m$).
Hence, in what follows the insignificant (in this region) exponential
dependence of $j_s(\rho ,0)$ will be neglected.
\begin{figure}[tbh]
\vspace{-20pt} \centerline{\epsfig{file=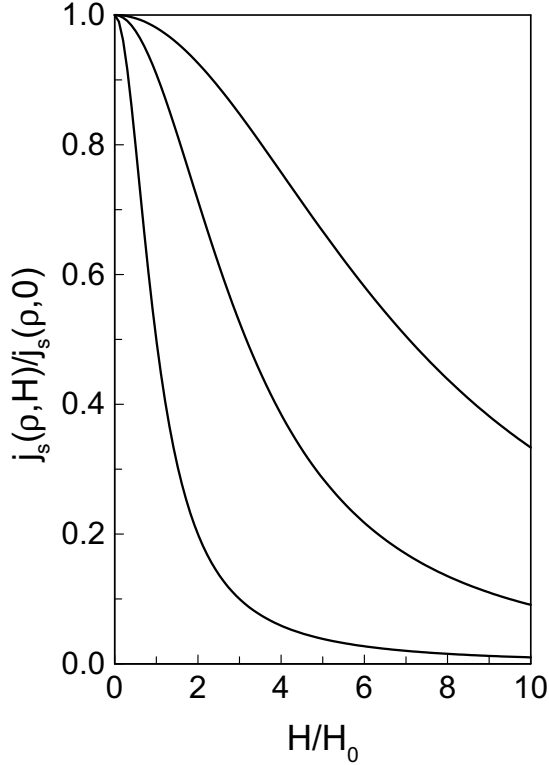,width=8cm}}
\vspace{-20pt} \caption{{\small {\sl Magnetic field dependence of the
dislocation-induced critical current density for various values of
the number of screw dislocations (increasing from bottom to top):
$n\equiv \rho /\rho _0=0.1;0.5;0.9$.}}} \vspace{10pt}
\end{figure}
\par
Fig.1 shows the behavior of the dislocation-induced critical current
density $j_s(\rho ,H)/j_s(\rho ,0)$, calculated according to Eqs.(7),
(12), and (13), versus applied magnetic field $H/H_0$ for various
number of dislocations $n\equiv \rho /\rho _0$. Notice that at least
in a qualitative agreement with the experimental observations in
screw dislocated single crystals~\cite{17}, Eqs.(7) and (13) suggest
a linear (for small $\rho $) increase of the critical current density
as well as a rather strong weakening of its field dependence (via the
increase of the characteristic Josephson field, $H_\rho $) with the
number of defects. It is worthwhile to mention that in view of the
normalization condition $\int_0^Ldx\rho (x)=b$ and Eq.(11), the
length of the JJ contact $L$ will be altered by external forces $\rho
_0$, namely $L=3b/2\rho _0$. In turn, due to the equilibrium equation
(8), it means that within our scenario the number of dislocations
$\rho $ does not depend on external forces and is considered as a
constant parameter. To estimate $\rho _0,$ let us consider the
pinning mechanism due to the screw dislocations. According to
McElfresh et al.~\cite{45} (who treated the correlation of surface
topography and flux pinning in YBCO thin films), the strain field of
a dislocation can provide a mechanism by which the superconducting
order parameter can be reduced, making it a possible site for core
pinning. The elastic pinning mechanism could also be responsible for
the large pinning forces associated with dislocation defects. There
are two types of elastic pinning mechanisms possible, a first-order
(parelastic) interaction and a second-order (dielastic) interaction.
In the case of a screw plane defect, the parelastic pinning can be
shown to be negligible~\cite{45}. However, the dielastic interaction
comes about because the self-energy of a defect depends on the
elastic constants of the material in which it forms. Since the
crystalline material in a vortex core is stiffer, there is a higher
energy bound up in the defect. For a tetragonal system with a screw
dislocation, the energy density due to the defect strain is $(1/2)\mu
\epsilon _{44}^2$, where $\mu $ is the shear
modulus and $\epsilon _{44}$ is the shear strain. For a screw plane, $%
\epsilon _{44}\cong b/r$, where $r$ is the distance from the center
of the defect and $b$ is the Burgers vector of the defect. The
interaction energy density between the vortex and the screw plane is
just $(1/2)\Delta \mu \epsilon _{44}^2$, where $\Delta \mu $ is the
difference in shear modulus between superconducting and normal
regions. For a vortex a distance $r$ away from the defect, the
interaction energy (per unit length of dislocation) reads
\begin{equation}
{\cal E}_d(r)=\frac 12\Delta \mu \epsilon _{44}^2(\pi \xi _0^2)
\end{equation}
Setting $r\simeq \xi _0$, we get finally for the elastic force
affecting the distribution of dislocations inside a crystal
\begin{equation}
f_{el}=\Delta \mu \left( \frac{b^2}{2\xi _0}\right) ,
\end{equation}
which in view of Eqs.(10)-(12) results in $\rho _0=2f_{el}/\mu b=$ $%
(b/\xi _0)(\Delta \mu /\mu )$. Taking the reported~\cite{45} for YBCO
values of $\mu \simeq 10^{10}N/m^2,$ $\Delta \mu =10^{-5}\mu $, and
$b\simeq \xi _0\simeq 1$ $nm$, we find that $f_{el}\simeq 5\times
10^{-5}N/m$ and $\rho _0\simeq 10^{-5}$. In turn, using $n_d\simeq
10^{13}m^{-2}$ for the maximum number of dislocations observed in
screw dislocated thin films~\cite{15,17,46}, which corresponds to our
dimensionless quantity $\rho _{\exp }\simeq \pi b^2n_d/4\simeq
8\times 10^{-6}$, we get $\rho _{\exp }/\rho _0\simeq 0.8$. On the
other hand, as it follows from Fig.1, for $H/H_0=5$ our model
predicts $j_s(\rho ,H)/j_s(\rho ,0)=0.8$, which according to Eqs. (7)
and (13) results in $\rho /\rho _0\simeq 0.9$, in a good agreement
with the observations. Furthermore, using the fact that in the
above-mentioned experiments~\cite{17} substantial pinning enhancement
due to screw dislocations has been observed for applied fields up to
$H=0.05$ T, we get $H_0\simeq 0.01$ T for an estimate of the
characteristic intrinsic Josephson field. Using~\cite{45} $\lambda
_L\simeq 150$ nm for the zero-temperature London penetration depth in
YBCO crystals, the above value of $H_0$ brings about the estimate of
the length of the dislocation-induced intrinsic JJ contact $L\simeq
100nm$, that is $t\ll L<\lambda _J$ (where $t\simeq \xi _0$ is the TB
insulator thickness and the Josephson penetration depth $\lambda
_J\simeq 1\mu m$ for the intragrain critical current density $j_{c0}
\simeq 10^{10}A/m^2$) in a satisfactory agreement with the
applicability conditions for both the thin TB model~\cite{9} and the
model of small Josephson junctions~\cite{39}. Moreover, since the
vortex lattice parameter at $B=0.05T$ is $a=\sqrt{\phi _0/B}\simeq
100 nm$, a dislocation-induced Josephson contact (of length $L\simeq
100nm$) will indeed provide the optimum pinning center for applied
fields up to $B=0.05T$~\cite{17}. It is worthwhile to mention that a
pinning force per unit length $f_p=10^{-4}N/m$ (which is of the same
order of magnitude as the above-considered elastic force $f_{el}$
acting upon dislocations), calculated in the single vortex limit,
$f_p=j_{c0}\phi _0$, corresponds to a critical current density of
$j_{c0}=5\times 10^{10}A/m^2$, in reasonable agreement with the
$j_{c0}$ found in dislocated YBCO thin films~\cite{17,45}.

\subsection{Electric field effects}

Let us consider now within our scenario a rather interesting
phenomena observed in screw dislocated HTS thin films in the applied
electric field~\cite{18,47}. Since in ionic crystals dislocations can
accumulate (or trap, due to the electric field around their cores)
electrical charge along their length, application of an external
electric field can sweep up these additional carriers, refreshing the
dislocation cores, that is releasing the dislocation from a cloud of
point charges (which, in field-free configuration, screen an electric
field of the dislocation line for electrical neutrality). As is
well-known~\cite{50,51,52} the charges on dislocations play a rather
important role in charge transfer in ionic crystals. If a dislocation
is oriented so that it has an excess of ions of one sign along its
core, or if some ions of predominantly one sign are added to or
removed from the end of half-plane, the dislocation will be charged.
In thermal equilibrium it would be surrounded by a cloud of point
defects of the opposite sign to maintain electrical neutrality. A
screw dislocation can transport charge normal to the Burgers vector
if it can carry vacancies with it. As Eshelby~\cite{52} pointed out,
a dislocation resembles a surface in that it also can act as a source
or sink of vacancies. Utilizing the ionic (perovskite-like) nature of
the HTS crystals~\cite{53}, we propose a possible mechanism of the
critical current enhancement in screw dislocated YBCO thin
films~\cite{18} based on the electric-field induced converse
piezoeffect due to the high polarizability of the defected medium in
these materials. Piezoelectric effect, that is a transverse
polarization induced by dislocation motion in ionic crystals, is
certainly cannot be responsible for large field-induced effects in
YBCO thin films because of too low rate of dislocation motion at the
temperatures used in these experiments (another possibility to
observe the direct piezoeffect in dislocated crystals is to apply an
external stress field~\cite{50}). On the contrary, the converse
piezoeffect, i.e., a change of dislocation-induced strain field in
applied electric field can produce a rather considerable change of
the critical current density in screw dislocated HTS thin films.
Indeed, in applied electric field $\vec E$ the dislocation will be
influenced by the external force $f_e(E)=b\sigma _e(E)$, where
$\sigma _e(E)=\mu \epsilon _{44}(E)$ and the change of the
dislocation-induced shear strain field in the applied electric field
is defined as follows
\begin{equation}
\epsilon _{44}(E)=\epsilon _0\epsilon _rdE\cos \theta
\end{equation}
Here $d$ is the absolute value of the converse piezoeffect coefficient, $%
\theta $ stands for the angle between the screw dislocation line and
the direction of an applied electric field, $\epsilon _r$ is the
static permittivity which accounts for the long-range polarization of
dislocated crystal (see below), and $\epsilon _0=8.85\times
10^{-12}F/m$. According to Whitworth~\cite{50}, the converse
piezoeffect coefficient $d$ in ionic dislocated
crystals can be presented in the form
\begin{equation}
\frac 1d=\frac{q\rho }b
\end{equation}
Here $\rho $ is the (dimensionless) dislocation density, and
$q=e^{*}/w$ is
the effective electron charge per unit length of dislocation $w$ (in fact, $%
w $ coincides with the thickness of the YBCO films~\cite{18}).
Finally, due to Eq.(12) with $\rho _0(E)=(1+2f_e(E)/\mu b)$ and
$f_e(E)$ given by Eqs.(16) and (17), electric-field induced change of
the critical current density
reads 
\begin{equation}
j_s(\rho ,\vec E)=2j_c\frac \rho {\rho _0^2(\vec E)}=j_s(\rho
,0)\left( 1+ \frac{\mid {\vec E}\mid }{E_0(\rho )}\cos \theta \right)
^{-2},
\end{equation}
where 
\begin{equation}
j_s(\rho ,0)=2j_c\frac \rho {\rho _0^2(0)},\qquad E_0(\rho )=\frac{q\rho }{%
\epsilon _0\epsilon _rb}.
\end{equation}
Here, $\rho _0(0)\equiv \rho _0(E=0)=\Delta \mu /\mu $.

To attribute the above-mentioned mechanism to the field-induced
changes of the critical current densities in dislocated YBCO thin
films, we propose the following scenario~\cite{s5}. Depending on the
gate polarity, a strong applied electric field will result either in
trapping the additional point charges by dislocation cores (when
$\cos \theta =+1$) reducing the pinning force density, or in sweeping
these point charges (surrounding the dislocation line to neutralize
the net charge) up of the dislocation core (for the opposite polarity
when $\cos \theta =-1$), thus increasing the core vortex pinning by
fresh dislocation line. Using the fact that the thickness $w$ of YBCO
thin films in the experiments carried out by Mannhart et
al.~\cite{18} was $\simeq $ $70\dot A$, for the linear charge per
length of dislocation we get the value $q=e^{*}/w=2\times
10^{-11}C/m$, that is $qb\simeq 0.1e^{*}$. To get an estimate of the
threshold field $E_0(\rho )$, let us discuss the relationship between
the above-introduced static permittivity of the dislocated crystal
$\epsilon _r$ and the more recognized dielectric constant which in
YBCO\ crystals is found~\cite{53} to be $\epsilon _{YBCO}\simeq 25$.
Let $S_{YBCO}$ be the crystal area, $N_d$ the number of dislocations,
and $S_d$ the area occupied by a single dislocation line ($S_d\simeq
\pi b^2$ where $b$ is the magnitude of the Burgers vector). Then,
approximately, $\epsilon _r/\epsilon _{YBCO}\simeq S_dN_d/S_{YBCO}$,
or $\epsilon _r\simeq 4\rho \epsilon _{YBCO}$, where $\rho \simeq \pi
b^2N_d/4S_{YBCO}$ is the (dimensionless) dislocation density.
Finally, using the maximum value of the dislocation density $n\simeq
4\rho /\pi b^2\simeq 10^9cm^{-2}$ observed in screw dislocated YBCO
thin films~\cite{15,17,46}, for an estimate of the threshold field
$E_0(\rho )$ Eq.(19) predicts the value of $5\times 10^5V/cm$ which
corresponds to the gate voltage $V_G=25V$ and reasonably agrees with
the typical values of the breakdown voltages used in the
above-mentioned experiments. For self-consistency of the
above-proposed scenario, it is important to mention that the induced
(by applied electric field) polarization force $f_e$ is compatible to
the elementary pinning force $f_p$, introduced in the previous
Section. Indeed, using the above-discussed parameters, we get from
Eqs.(16)-(19) that for gate voltage $V_G=25V$ and $\cos \theta =1$, $%
f_e=enV_G/4\simeq 10^{-5}N/m$. Let us briefly consider the influence
of the applied magnetic field on the electric-field mediated converse
piezoeffect. In view of Eqs.(13) and (16), external electric field
will result in the following change of the characteristic
field 
\begin{equation}
\frac{H_\rho (\vec E)}{H_\rho (0)}\simeq 1-\left( \frac{\Delta H_\rho (0)}{%
H_0}\right) \left( \frac{\mid {\vec E}\mid }{E_0(\rho )}\right) \cos
\theta ,
\end{equation}
where 
\begin{equation}
\Delta H_\rho (0)\equiv H_\rho (0)-H_0,
\end{equation}
and the characteristic fields $H_\rho (0)$ and $H_0$ are defined by
Eqs.(13) and (8), respectively. As it follows from Eqs.(7), and
(18)-(21), depending on the gate polarity the magnetic field
dependence of the critical current density in dislocated crystals
will be changed in a different way, in agreement with
observations~\cite{18}. It is also worthwhile to mention that since
recently a special attention has been given to the so-called electric
field effects (FEs) in JJs and granular
superconductors~\cite{4a,5a,6a,7a,8a,9a}. The unusually strong FEs
observed in bulk HTS ceramics~\cite{4a} (including a substantial
enhancement of the critical current, reaching $\Delta
I_c(E)/I_c(0)=100\%$ for $E=10^7V/m$) have been attributed to a
crucial modification of the original weak-links structure under the
influence of very strong electric fields. This hypothesis has been
corroborated by further investigations, both experimental (through
observation of the correlation between the critical current behavior
and type of weak links~\cite{5a}) and theoretical (by studying the
FEs in $SNS$-type structures~\cite{6a} and $d$-wave granular
superconductors~\cite{7a}). Among other interesting field induced
effects, one can mention the FE-based Josephson transistor~\cite{8a},
Josephson analog of the {\it magnetoelectric effect}~\cite{9a}
(electric field generation of Josephson magnetic moment in zero
magnetic field), and recently predicted~\cite{10a} giant enhancement
of {\it nonlinear} (i.e. $\nabla T$ - dependent) thermal conductivity
$\kappa $, reaching $\Delta \kappa (T,E)/\kappa (T,0)=500\%$ for
relatively low (in comparison with the fields needed to observe a
critical current enhancement~\cite{4a,5a}) applied electric fields
$E$ matching an intrinsic thermoelectric field $E_T=S_0|\nabla T|$,
where $S_0$ is the Seebeck coefficient.

\section{Fishtail anomaly (Peak effect)}
\subsection{Stoichiometric melt-textured crystals}

Let us apply our scenario to discuss a possible origin of the
so-called fishtail anomaly of the critical current density vs.
applied magnetic field in stoichiometric melt-textured crystals
possessing a large number of dislocations~\cite{25,26}. As is
well-known~\cite{9,41}, any external force leading to the
redistribution of dislocations in the crystal, change the volume
content of the martensite (superconducting) phase in HTS sample.
Under an influence of external magnetic field, for instance, the
dislocation will be affected by additional stress fields, $\sigma
_m$, due to the differences between the magnetic susceptibilities of
two phases. When one of these phases falls into the superconducting
state, its magnetic properties will change drastically in the
external magnetic field. If $B_n$ and $B_s$ are the magnetic
inductions of the defect (normal) and superconducting regions,
respectively, then such a dislocation will be influenced, in nonzero
magnetic field, by a force (Cf. the electric-field induced force
$f_e$ from previous Section): 
\begin{equation}
f_m=b\sigma _m=\frac b2\Delta BH.
\end{equation}
Here $\Delta B=B_n-B_s$,$B_n=\mu _0H$, $B_s=\mu _0H+4\pi M(H)$, and
$\mu _0=4\pi \times 10^{-7}N/A^2$. To choose an appropriate form of
the vortex lattice magnetization $M(H)$ sensitive to the magnetic
anomalies at the region of $B\simeq 1T$, we note that according to
our scenario, the strained regions near dislocation cores result in a
lower $H_{c2}^{*}(T)$ as compared to the YBCO matrix $H_{c2}(T)$.
Indeed, the order parameter $\Delta _{TB}$ near the TB is supposed to
be reduced as compared to its bulk value $\Delta _0$ as follows,
$\Delta _{TB}=\Delta _0(b/\xi _n)$, where $\xi _n$ is the decay
length of the TB-induced Josephson contact. In turn, this leads to a
similar depression of the superconducting coherence length $\xi
_{TB}=\xi _0(\xi _n/b)$ and the upper critical field,
$H_{c2}^{*}(T)\propto \phi _0/\xi _{TB}^2=H_{c2}(T)(b/\xi _n)^2$,
where $H_{c2}(T)=H_{c2}(0)(1-T^2/T_c^2)$, $H_{c2}(0)\propto \phi
_0/\xi _{0}^2$, and $\xi _0=\hbar v_F/\pi \Delta _0$. Since for YBCO
crystals, $\xi _0\simeq 1nm$, $\mu _0H_{c2}(0)\simeq 140T$, and
$T_c\simeq 90K$, we get $H_{c2}^{*}(T)\simeq 2T$ for $T=70K$. Thus,
with a good accuracy we can use a high-field limit for vortex lattice
magnetization (valid for $H\le H_{c2}^{*}$), $4\pi M(H)=-\mu
_0(H_{c2}^{*}-H)/2\kappa ^{*2}$ where $\kappa ^{*}=\kappa _0(b/\xi
_n)^2$ is the Ginzburg-Landau parameter near the TB (recall that for
YBCO $\kappa _0\simeq 90$). According to Eqs.(7) and (13),
dislocation-induced critical current density $j_s(\rho ,H)$, driven
by the
magnetic-type external force (22), will exhibit a maximum at some field $%
H^{*}(\rho )$ which is defined as the solution of the balance
equation $\rho _0(H^{*})=\rho $, where $\rho _0(H)=2f_m(H)/\mu b= \mu
_0H(H_{c2}^{*}-H)/2\mu \kappa ^{*2}$. In view of the above-given
explicit form of $\rho _0(H)$, the balance equation produces the
following (dislocation-dependent) threshold field
\begin{equation}
H^{*}(\rho )=H_{c2}^{*}\left [1- \sqrt{1-\left
(\frac{H_d}{H_{c2}^{*}}\right )^2}\right ]
\end{equation}
where $H_d=\sqrt{8\mu \rho \kappa ^{*2}/\mu _0}$.
\begin{figure}[tbh]
\vspace{-10pt} \centerline{\epsfig{file=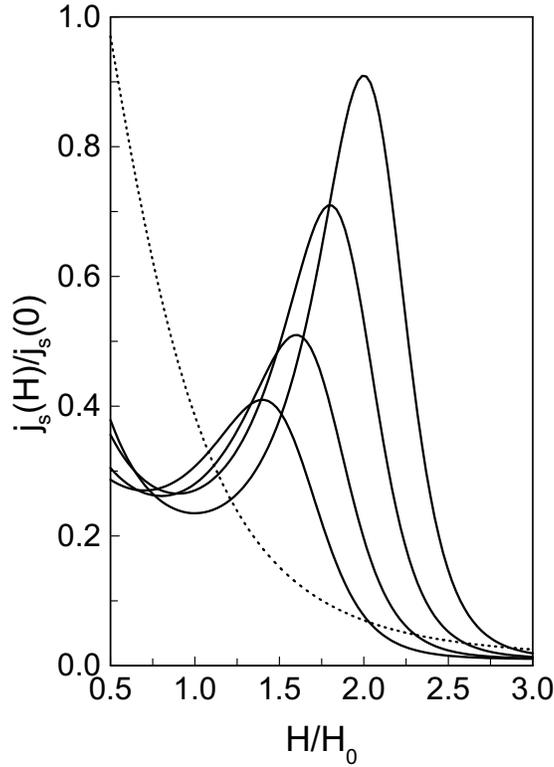,width=8cm}}
\vspace{-20pt} \caption{{\small {\sl The behavior of the normalized
critical current density $j_s(H)/j_s(0)$ of melt-textured sample
versus reduced magnetic field $H/H_0$ (peak effect) for various
values of the number of dislocations (increasing from bottom to top):
$n\equiv \rho /\rho _0=0.2;0.5;0.7;0.9$. The dotted line corresponds
to defect-free sample ($\rho =0$).}}} \vspace{10pt}
\end{figure}
\par
The existence of distinct magnetic properties along both sides of the
twinning boundary leads, in nonzero magnetic field, to the growing of
this twin through the crystal. This process can be stopped by the
friction of dislocation via the crystalline lattice or by generation
of an excess nonequilibrium concentration of (oxygen) vacancies (see
the next Section). When the applied magnetic field reaches the
threshold field $H^{*}(\rho )$, the moving dislocation is blocked,
thus leading to the adjustment of the defect structure to the
penetrating vortex lines and, as a result, to the enhancement of the
pinning forces and the recovery of the critical current density. When
magnetic field overcomes this barrier, the pinning becomes less
effective, and a rather smooth fall-off of the critical current is
restored.  Figure 2 depicts the predicted behavior of the normalized
critical current density, $j_s(H)/j_s(0)$, of melt-textured sample
versus reduced magnetic field $H/H_0$, calculated according to
Eqs.(7), (12) and (13), for various values of the number of
dislocations. Using $B^{*}=\mu _0H^{*}\simeq 1T$ for the typical
field region where anomalous fishtail behavior of critical currents
has been observed~\cite{25,26}, we can estimate a number of
dislocations $n_{MT}$ needed to produce a peak-effect in
stoichiometric melt-textured samples. Namely, taking $b\simeq 1nm$,
$\mu \simeq 10^{10}N/m^2$, $\mu _0H_{c2}^{*}(0)\simeq 5T$,  and
$\kappa ^{*}\simeq 4$, we get $\mu _0H_d\simeq 1T$ and $n_{MT}\simeq
4\rho /\pi b^2\simeq 2\times 10^8cm^{-2}$ which is about $50\%$ of
the maximum number of dislocations found in the melt-textured
materials~\cite{25,26}. Finally, using the above parameters, we can
estimate the magnitude of the magnetic force, $f_m$. The result for
$B\simeq 1T$ is $f_m\simeq 2\times 10^{-5}N/m$ which is quite
comparable with the elementary pinning force density $f_p$, elastic
force $f_{el}$, and electric field polarization force $f_e$.

\subsection{Deoxygenated single crystals}

To apply the above-considered scenario for the description of the
anomalous magnetic behavior observed in the oxygen-deficient single
crystals, we should take into account existence of a rather important
force of inelastic origin (so-called osmotic force $f_0$) which,
together with the above-mentioned magnetic force, $f_m$, will define
the ultimate defect structure of these crystals. The origin of this
force in deoxygenated crystals is due to an excess nonequilibrium
concentration of oxygen vacancies $c_v$ (with $c_v\neq 1$). Unlike
the ordinary oxygen diffusion $D=D_0e^{-U_d/k_BT}$ in
$YBa_2Cu_3O_{7-\delta }$ which is extremely slow even near $T_c$ (due
to a rather high value of the activation energy $U_d$ in these
materials, typically $U_d\simeq 1eV$), the osmotic (pumping)
mechanism~\cite{57} can substantially facilitate oxygen transport in
underdoped crystals (with oxygen-induced dislocations). This
mechanism relates a local value of the chemical potential (chemical
pressure) $\mu ({\bf x})=\mu (0)+\nabla \mu \cdot {\bf x}$ with a
local concentration of point defects as follows $c({\bf x})=e^{-\mu
({\bf x})/k_BT}$. Indeed, when in such a crystal there exists a
nonequilibrium concentration of vacancies, dislocation is moved for
atomic distance $b$ by adding excess vacancies to the extraplane edge
and as a result the force $f_o$ produces the work $bf_o$. The
produced work is simply equal to the chemical potential of added
vacancies, so that $bf_o=\mu _v/b$. What is important, this mechanism
allows us to explicitly incorporate the oxygen deficiency parameter
$\delta $ into our model by relating it to the excess oxygen
concentration of vacancies $c_v\equiv c(0)$ as follows $\delta=1-c_v$
(in most interesting cases $\delta \ll 1$). As a result, the chemical
potential of the single vacancy reads $\mu _v\equiv \mu (0)=-k_BT\log
(1-\delta )$, leading to the following explicit form of the osmotic
force~\cite{54,55,56}
\begin{equation}
f_0=-\frac{k_BT}{b^2}\log (1-\delta )\simeq \frac{k_BT}{b^2}\delta .
\end{equation}
Remarkably, the same osmotic mechanism was used by Gurevich and
Pashitskii~\cite{b4} to discuss the modification of oxygen vacancies
concentration in the presence of the TB strain field. Let us consider
influence of two competitive forces, $f_m$ and $f_0$, on the magnetic
field critical current behavior. In accordance with Eqs.(9), (12),
and (22), the external force dependent part of the dislocation density, $%
\rho _0$, reads
\begin{equation}
\rho _0(\delta ,H)=\rho _0(\delta ,0)\left[ 1+\frac{H(H_{c2}^{*}-H)}{H_\delta ^2}%
\right ],
\end{equation}
where
\begin{equation}
\rho _0(\delta ,0)\equiv \frac{2k_BT}{b^3\mu }\delta ,\qquad H_\delta
\equiv \sqrt{\frac{4k_BT\kappa ^{*2}\delta }{b^3\mu _0}.}
\end{equation}
Thus, the non-zero oxygen deficiency (which is responsible for the
sample nonstoichiometry) results in an intrinsic characteristic
field, $H_\delta $, which reflects a subtle balance between extended
(dislocations) and point (oxygen vacancies) defects in a given
sample. In view of Eqs.(7) and (13), the above $\rho _0(\delta ,H)$
brings about the following explicit dependence of the
dislocation-induced critical current density
\begin{equation}
j_s(H)=2j_{c0}(\delta )\frac{\rho }{\rho _0^2(\delta ,H)}\left[
1+\left( \frac H{H_0}\right) ^2\left( 1-\left( \frac \rho {\rho
_0(\delta ,H)}\right) ^2\right) ^2\right] ^{-1}
\end{equation}
\begin{figure}[tbh]
\vspace{-20pt} \centerline{\epsfig{file=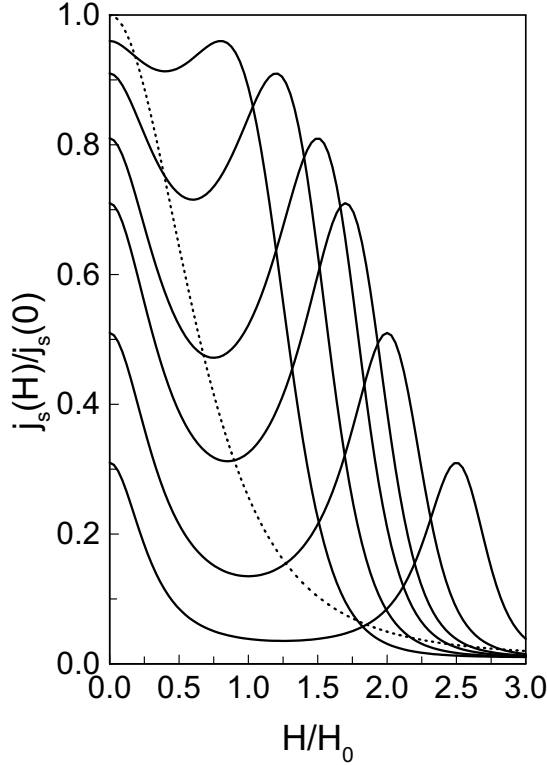,width=8cm}}
\vspace{-20pt} \caption{{\small {\sl The behavior of the normalized
critical current density $j_s(H)/j_s(0)$ versus reduced magnetic
field $H/H_0$ for various values of the oxygen deficiency parameter
(increasing from top to bottom): $\delta =0.01; 0.05; 0.07; 0.1;
0.15; 0.2$. The dotted line corresponds to stoichiometric sample
($\delta =0$).}}} \vspace{10pt}
\end{figure}
\par
Here we have taken into account that according to the
observations~\cite{23} $j_{c0}(\delta )\simeq j_{c0}(0)(1-\delta )$.
Fig.3 illustrates the evolution of $j_s(H)/j_s(0)$ versus applied
magnetic field $H/H_0$ with increasing the oxygen deficiency $\delta
$ (for $\rho =\rho _0(\delta ,0)$). As is seen, below some threshold
concentration of oxygen defects, $\delta _c$, critical current
density exhibits a peak-like behavior. The peak is shifted to lower
fields with increasing the oxygen deficiency and disappears at
$\delta =\delta _c$, in agreement with the observations. Due to
Eqs.(25)-(27), $j_s(H)$ passes through the maximum above some
threshold field, $H^{*}(\delta ,T)$, which is defined via the balance
equation $\rho _0(\delta ,H^{*})=\rho $. Taking into account the
explicit form of $\rho _0(\delta ,H)$ (see Eqs.(25)), the threshold
field reads
\begin{equation}
H^{*}(\delta ,T)=H_{c2}^{*}\left [1- \sqrt{1-\left
(\frac{H_d}{H_{c2}^{*}}\right )^2\left ( 1-\frac{\delta}{\delta
_c}\right )}\right ],
\end{equation}
or alternatively
\begin{equation}
H^{*}(\delta ,T)=H_{c2}^{*}\left [1- \sqrt{1-\left
(\frac{H_d}{H_{c2}^{*}}\right )^2\left ( 1-\frac{T}{T_0}\right
)}\right ],
\end{equation}
where 
\begin{equation}
\delta _c\equiv \frac{\mu b^3\rho }{2k_BT},\qquad T_0\equiv \frac{\mu
b^3\rho }{2k_B\delta }.
\end{equation}
Thus, in addition to the strong $\delta $-dependence, Eq.(28)
predicts decrease of the peak-effect threshold field, $H^{*}(\delta
,T)$, with increasing the temperature as well, also in agreement with
the observations~\cite{19,20,21,22,23}. Taking $b\simeq 1nm$, $\delta
_c=0.23$, $\mu _0H_d\simeq 1T$, and $\mu _0H_{c2}^{*}(T=70K)\simeq
2T$, for $\delta =0.05$ and $T=70K$ we obtain the estimate of the
threshold field $B^{*}=\mu _0H^{*}\simeq 1T$, which is very close to
the fields where the anomalous phenomena in $YBa_2Cu_3O_{7-\delta }$
single crystals have been observed~\cite{20}. According to Eq.(24),
the above estimates predict $f_o\simeq 5\times 10^{-5}N/m$ for the
osmotic force magnitude. At the same time,
using the experimentally found~\cite{22} critical value of oxygen deficiency $%
\delta _c=0.23$, Eq.(30) suggests an estimate of the number of
dislocations, $n_{DO}$, needed to produce the above-discussed
critical current enhancement (fishtail anomaly) in deoxygenated HTS
single
crystals. Namely, for $T=70K$ and $\mu =10^{10}N/m^2$, we get $%
n_{DO}\simeq 4\rho /\pi b^2\approx 2\times 10^6cm^{-2}$. Notice that
this is only $1\%$ of the number of defects $n_{MT}$ needed to
produce the fishtail anomaly in well-oxygenated melt-textured
crystals (see the previous Section). It means, in fact, that in
contrast to the melt-textured case, in oxygen-depleted single
crystals dislocations play an auxiliary (as compared to the points
defects, oxygen vacancies), yet important role, providing a
background for oxygen defects redistribution inside a crystal and
thus adjusting the pinning ability of the material. A rather strong
correlation
between dislocations and oxygen defects is seen also through $\rho $%
-dependence of the critical oxygen deficiency parameter $\delta _c$
(see Eq.(30)). The latter parameter has a simple physical meaning.
Namely~\cite{22}, for $\delta >\delta _c$ oxygen-rich superconducting
grains are separated by oxygen-poor insulating boundaries so that
there is no superconducting path through the sample. Notice also (by
comparing the low field regions in Fig.3 for different $\delta $)
that at a given number of dislocations, increase of oxygen defects
leads to a more flat behavior of the critical current at small fields
indicating that point defects are the major pinning centers in this
region~\cite{19}. In a similar way, $T_0$ plays a role of the
phase-locking temperature above which the superconductivity between
oxygen-rich grains is suppressed and the fishtail phenomenon
disappears. For the above-mentioned parameters this characteristic
temperature is estimated to be $T_0\simeq 87K$.  It is interesting
also to mention the experiments on electromigration of oxygen in HTS
single crystals conducted by Moeckly et al.~\cite{59}. They have
observed that the superconducting properties of the disordered
regions caused by electromigration are indicative of a filamentary
superconductive system shunted by nonsuperconductive Ohmic parts. If
we assume, after Moeckly et al.~\cite{59}, that the so-called
electron-wind contribution dominates the phenomenon, the
electromigration force acting on the oxygen ions reads
\begin{equation}
f_{em}=\frac{2mv_F\sigma }{be}J_{em}.
\end{equation}
Here $J_{em}$ is the applied critical current density, $v_F$ is the
Fermi velocity, and $\sigma $ is the oxygen ion scattering cross
section. Taking into account the values of the parameters used in the
above-mentioned experiments~\cite{59}, namely $J_{em}=5\times
10^6A/cm^2$, $v_F=2\times 10^7cm/s $, and $\sigma \approx b^2\approx
10^{-14}cm^2$, we get for an estimate of the electromigration force
$f_{em}\simeq 10^{-4}N/m$. Remarkably, this force is of the same
order of magnitude as the above-considered magnetic and osmotic
forces. Besides, the described in this Section scenario has been
used~\cite{y1} to explain the observed~\cite{y2} irreversibility line
$T_{c}(H)$ crossover (between Almeida-Thouless and Gabay-Toulouse
behavior) in deoxygenated HTS. And finally, a few novel interesting
effects in intrinsically granular non-stoichiometric material have
been recently predicted~\cite{s2}, including Josephson chemomagnetism
(chemically induced magnetic moment in zero applied magnetic field)
and its influence on a low-field magnetization (chemically induced
paramagnetic Meissner effect), and magnetoconcentration effect
(creation of extra oxygen vacancies in applied magnetic field) and
its influence on a high-field magnetization (chemically induced
analog of fishtail anomaly).

\vspace{8mm} \section{Irradiated single crystals} \vspace{5mm}

There are two possibilities to adapt our scenario for a qualitative
description of the anomalous magnetic phenomena observed in
irradiated HTS single crystals. We can either consider the
well-aligned tracks of damaged material, produced by particle
irradiation, as potential sources of intragrain weak links (provided
that the track's diameter is of the order of the coherence
length~\cite{56}) or, alternatively, exploit the
well-known~\cite{60,61} fact that irradiation produces a large
density of extended defects (in particular, dislocations) in these
materials. The experiments revealed that there is some threshold
magnetic field, $H^{*}(\Phi )$, which is almost linearly increases
with the fluence $\Phi $ and above which a substantial increase of
the critical current density, $j_s(\Phi ,H)$, has been
observed~\cite{29,30}. Besides, it was found that the magnitude of
fluence, $\Phi ^{*}(H)$, which optimizes $j_s(\Phi ,H)$, increases
almost linearly with the applied field. As it was observed~\cite{31},
the strain and structural disorder of the amorphous region propagates
into the crystal lattice in a direction normal to the ion track. It
means that, for an isolated amorphous track, a radially symmetric
displacement field $u(r)\simeq E_{eff}(R^2/r)$ (where $R$ is the
radius of the amorphous track, and for YBCO $E_{eff}\simeq 0.02$)
occurs in the matrix around the ion track (Cf. the appearance of a
displacement field around a screw dislocation). According to Weaver
et al.~\cite{61}, irradiation of fluence $\Phi $ induces the network
of dislocations of density $n(\Phi )=n(0)+n_m(1-\exp (-\alpha \Phi
))$, where $n(0)$ is the dislocation density of a virgin
(pre-irradiated) sample, $n_m$ the maximum density after irradiation,
and $\alpha $ is the so-called displacement cross section which is
related to the number of displaced atoms per particle per unit
length. Assuming that $\alpha \Phi \ll 1$ (which is usually the case
for real experiments), in the dimensionless units the above equation reads $%
\rho (\Phi )\simeq \rho (0)+\alpha \Phi $, where $\rho (\Phi )\equiv
n(\Phi )/n_m$ and $\rho (0)\equiv n(0)/n_m$. Let us consider the
magnetic field behavior of the critical current density, $j_s(\Phi
,H)$, upon irradiation. According to our scenario for
weak-links-mediated critical current
enhancement in defected samples, the irradiation-induced behavior of $%
j_s(\Phi ,H)$ will show the same peculiarities we have discussed
earlier considering the fishtail anomaly in melt-textured and
deoxygenated crystals. Indeed, in view of the fluence dependence of
the characteristic field $H_\rho $ (through $\rho (\Phi )$, see
Eq.(13)), it is easy to verify that $j_s(\Phi ,H)$ will have a
maximum at some threshold field, $H^{*}(\delta ,\Phi )$, which is
defined (by analogy with the threshold field for oxygen-deficient
sample, Cf. Eq.(28)) as the solution of the balance equation $\rho
_0(H^{*},\delta )=\rho (\Phi )$. Taking into account the definitions
of $\rho _0(\delta ,H),$ given by Eqs.(25) and (26), and $\rho (\Phi
)$, the above balance equation results in the following
threshold field 
\begin{equation}
H^{*}(\delta ,\Phi )= H_{c2}^{*}\left [1- \sqrt{1-\left
(\frac{H_d(\delta )}{H_{c2}^{*}}\right )^2\left ( 1+\frac{\Phi}{\Phi
^{*}(\delta ,0)}\right )}\right ],
\end{equation}
where 
\begin{equation}
\Phi ^{*}(\delta ,0)=(\delta _c-\delta )\Phi _0^{*}, \qquad \Phi
_0^{*}=\frac{2k_BT}{\alpha \mu b^3}.
\end{equation}
Here $H_d(\delta )=H_d\sqrt{1-\delta /\delta _c}$ with $\delta _c$
and $H_d$ still governed by Eqs.(28)-(30) but with $\rho \equiv \rho
(0)$. Alternatively, we can introduce an optimum (maximum) value of
the fluence, $\Phi ^{*}(\delta ,H)$, as the solution of the equation
$\rho _0(H,\delta )=\rho (\Phi ^{*})$. The result is as follows
\begin{equation}
\Phi ^{*}(\delta ,H)=\Phi ^{*}(\delta ,0)\left[ 1+\frac
{H(H_{c2}^{*}-H)}{H_d^2(\delta )} \right]
\end{equation}
\begin{figure}[tbh]
\vspace{-20pt} \centerline{\epsfig{file=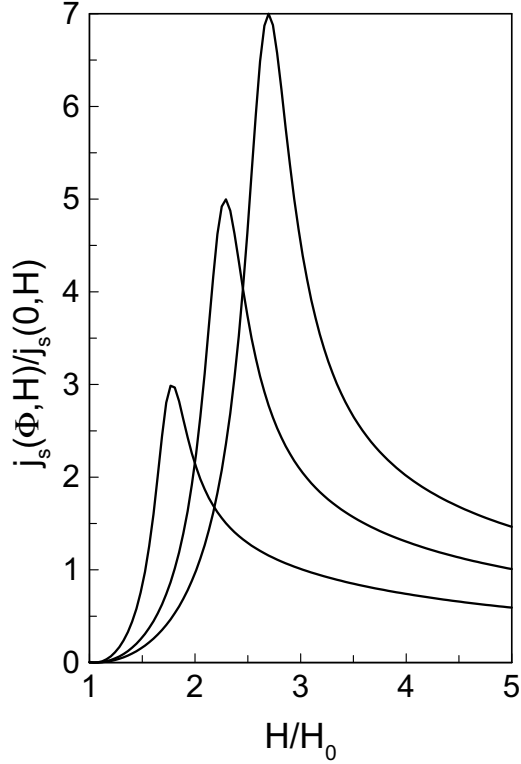,width=8cm}}
\vspace{-20pt} \caption{{\small {\sl The behavior of the normalized
critical current density $j_s(\Phi ,H)/j_s(0,H)$ versus reduced
magnetic field $H/H_0$ for various values of the reduced irradiation
fluence (increasing from bottom to top): $\Phi /\Phi _0=1;5;10$.}}}
\vspace{10pt}
\end{figure}
\par
Turning to the discussion of the above-obtained model predictions, we
notice that they cover practically all the peculiarities observed in
irradiated single crystals. Indeed, in view of Eqs.(7) and (13), the
irradiation-induced critical current density reads
\begin{equation}
j_s(\Phi ,H)=2j_{c0}(\Phi )\frac{\rho (\Phi )}{\rho _0^2(\delta
,H)}\left[ 1+\left( \frac H{H_0}\right) ^2\left( 1-\left( \frac{\rho
(\Phi )}{\rho _0(\delta ,H)}\right) ^2\right) ^2\right] ^{-1}
\end{equation}
Here we have taken into account the possibility of deterioration of
the pre-irradiated weak links (due to another defects) upon
irradiation. Within
the same accuracy ($\alpha \Phi \ll 1$), experiments~\cite{62} revealed that $%
j_{c0}(\Phi )\simeq j_{c0}(0)(1-\sqrt{\alpha \Phi }$). Let us
consider first the case of nearly fully oxygenated pre-irradiated
samples (with $\delta \simeq 0$). Fig.4 depicts the calculated
dependence of the critical current density $j_s(\Phi ,H)/j_s(0,H)$
upon applied magnetic field $H/H_0$ for various values of the
irradiation fluence $\Phi /\Phi _0$ (with $\Phi _0=\rho (0)/\alpha
$). As is seen, a pronounced peak appears which is shifted to higher
magnetic fields with increasing the irradiation fluence. And this
model prediction is also in agreement with the observations.
\begin{figure}[tbh]
\vspace{-20pt} \centerline{\epsfig{file=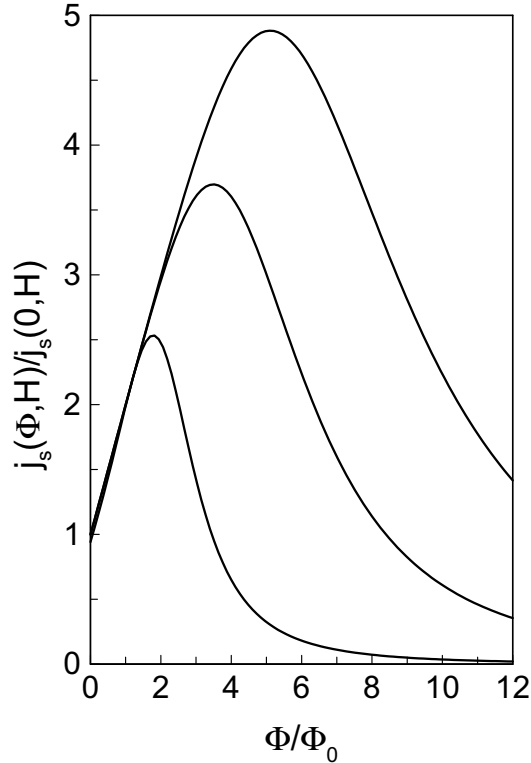,width=8cm}}
\vspace{-20pt} \caption{{\small {\sl The behavior of the normalized
critical current density $j_s(\Phi ,H)/j_s(0,H)$ versus reduced
irradiation fluence $\Phi /\Phi _0$,calculated according to Eq.(35),
for various values of the reduced magnetic field (increasing from
bottom to top): $H/H_0=1;2;3$.}}} \vspace{10pt}
\end{figure}
\par
In turn, Fig.5 shows the behavior of $j_s(\Phi ,H)/j_s(0,H)$ versus
irradiation fluence $\Phi /\Phi _0$ for different values of the
applied magnetic field $H/H_0$, calculated according to Eqs.(35) and
(25). Notice a tremendous increase of the critical current density
with fluence (especially at high magnetic fields where the
irradiation-induced defect structure is supposed to match optimally
the vortex lattice structure). The magnitude of the peak, $\Phi
^{*}(H)$, is seen to shift to higher fluences with increasing the
field, in agreement with observations~\cite{29,30}. Using the
experimental data of Hardy et al.~\cite{29,30} for the threshold
fields at different fluences, namely $B^{*}(\Phi =10^{11}cm^{-2})=1T$
and $B^{*}(\Phi =3\times 10^{11}cm^{-2})=2T$, we get from Eqs.(32)
and (33) $\Phi _0\simeq 5\times 10^{10}cm^{-2}$. Furthermore, using a
typical value of $\alpha \simeq 7\times 10^{-17}cm^2$~\cite{61}, we
can get an estimate of the defect number density in pre-irradiated
sample. The result is as follows, $n(0)\simeq 4\rho (0)/\pi b^2\simeq
2\times 10^8cm^{-2}$. Turning to the optimum value of the fluence,
$\Phi ^{*}(H)$, which has been measured during the same
experiments~\cite{29,30}, we deduce the following estimates: $\Phi
^{*}(B=1T)=5\times 10^{10}cm^{-2}$ and $\Phi ^{*}(B=2T)=3\times
10^{11}cm^{-2}$. In view of Eq.(34), the above values predict a
reasonable value for the intrinsic characteristic field
$B^{*}(0,0)\simeq 1T$. Let us briefly consider the case when
pre-irradiated sample is in a highly oxygen-deficient state (with
$\delta \simeq \delta _c$). Since $\Phi ^{*}(\delta ,0)=(\delta
_c-\delta )\Phi _0^{*}$, it means that a less value of the fluence is
required to be applied to the deoxygenated sample to get the same (as
for fully oxygenated sample) optimum properties of the critical
current density in irradiated sample. This conclusion is in agreement
with some recent experimental observations~\cite{31,37,37a,37b,37c}.
For example, Zhu et al.~\cite{31} have observed that the extent of
radiation damage strongly depends on the stoichiometry of the sample.
They found that an average diameter of the amorphous region for
$YBa_2Cu_3O_{7-\delta }$ crystals was $2R\simeq $7.8 nm, 5.3 nm, and
3.3 nm for $\delta \simeq 0.7$, $0.3$, and $0.01$, respectively.
Moreover, in oxygenated samples the irradiation-induced damage was
found~\cite{31} to be much less profound than that in
oxygen-deficient samples.

It is worthwhile to mention that a somewhat similar concept of
weak-links induced vortex pinning has been also considered by Lee et
al.~\cite{38}. Using a Monte-Carlo technique within a 3-D model of
Josephson junction arrays (JJAs), they found a substantial increase
of critical current densities due to the flux pinning by columnar
defects (for discussion of many other interesting effects in JJAs,
see, e.g., a recent review~\cite{s8}). Besides, the suggested in this
Section scenario can also be responsible for the observed~\cite{z1}
fluence induced shift in the irreversibility line (via the $\Phi $
dependence of the threshold field $H^{*}(\Phi )$)~\cite{56} and
allows to interpret the experimental data on $\Phi $ dependence of
the power-like current-voltage characteristics and non-Ohmic
resistive states in heavy-ion irradiated $YBa_{2}Cu_{3}O_{7-\delta }$
crystals~\cite{x1,56}.

\section{Conclusion}

In summary, a unified approach for description of the critical
current density improvement in dislocated, deoxygenated, and
irradiated superconductors was proposed based on intrinsic Josephson
effect which is supposed to be active in HTS materials. The model is
based on the existence of various competitive forces affecting a
rather delicate balance between extended (dislocations) and point
(oxygen vacancies) defects inside a crystal. The scenario
incorporates the idea of a quite strong correlation between the
intragrain weak links and vortex pinning, and implies that
practically any treatment of the superconducting sample (such as
sintering, melt-texturing, silver coating, thermal and mechanical
treatment, oxygenation/deoxygenation process, particle irradiation,
application of high magnetic and electric fields, etc) will
inevitably result in the rearrangement of the pre-treated defect
structure of the material to optimize its pinning ability as well as
its vital critical parameters (such as critical temperature, critical
fields, critical current density, magnetization, etc)~\cite{s7}.
Table I summarizes the external forces acting on dislocations
(including the equivalent elementary pinning force $f_p$), considered
in the present paper, as well as their estimates according to the
corresponding parameters deduced from the literature. Notice that all
of them are of the same order of magnitude supporting thus the idea
of unification of different physical phenomena related to the
critical current density improvement in high-temperature
superconductors.
\def\tstrutu{\vrule height3.7ex depth0ex
width0pt} \def\tstrutd{\vrule height 0ex depth1.2ex width0pt}
\def\tstrut{\vrule height3.7ex depth2.2ex width0pt} \def\m#1#2#3
{\multicolumn{#1}{#2}{#3}}
\begin{center}
\begin{table}[tbh]
\caption{{\small {\sl The external forces affecting the defect
structure arrangement within a sample: equivalent elementary pinning
force ($f_p$), elastic force ($f_{el}$), magnetic force ($f_m$),
electric force ($f_e$), osmotic force ($f_o$), electromigration force
($f_{em}$), and analog of irradiation force ($f_{\phi}$).}}}
\end{table}
\begin{tabular}{|c|c|c|} \hline
{{\bf Forces}}\tstrut  & {{\bf Parameters}}  \\[.3cm] \hline
 $f_{p}=j_{c}\phi _{0}\simeq 2\times 10^{-5}N/m$\tstrut
 & $j_{c}=10^{10}A/m^{2}$, $\phi _{0}=2\times 10^{-15}Wb$
 \\[.5cm]
 $f_{el}= \left( \frac{b}{2}\right )\Delta \mu \simeq 5\times 10^{-5}N/m$\tstrut
 & $\Delta \mu = 10^{5}N/m^{2}$, $b=1 nm$
 \\[.5cm]
 $f_{m}= \left( \frac{b}{2}\right )\Delta BH\simeq 3\times 10^{-5}N/m$\tstrut
 & $\Delta B\simeq \mu _{0}H \simeq 1T$, $b=1 nm$
 \\ [.5cm]
 $f_{e}=\left( \frac{1}{4}\right )enV_{G}\simeq 4\times 10^{-5}N/m$\tstrut
 & $V_{G}=25V$, $n= 10^{13}m^{-2}$  \\ [.5cm]
 $f_{o}=\left (\frac{k_{B}T}{b^2}\right )\delta \simeq 5\times 10^{-5}N/m$\tstrut
 & $\delta = 0.05$, $b=1 nm$, $T=70K$  \\ [.5cm]
 $f_{em}=\left (\frac{mv_{F}b}{e}\right )J_{em}\simeq 10^{-4}N/m$\tstrut
 & $J_{em}=10^{10}A/m^{2}$, $v_{F}\simeq 10^{5}m/s$
\\ [.5cm] $f_{\phi }\simeq\left (\frac{b\phi _0^2}{2\mu _0}\right )n_{\Phi}^2
 \simeq 2\times 10^{-5}N/m$\tstrut
 & $n_{\Phi} \simeq 10^{14}m^{-2}$, $b=1 nm$\\ [.5cm]\hline
\end{tabular}
\end{center}

\vspace{8mm} \leftline{\bf ACKNOWLEDGMENTS} \vspace{5mm}

Very useful discussions on the subject matter with Fernando
Araujo-Moreira, Marcel Ausloos, John Clem, Guy Deutscher, Mauro
Doria, Ted Geballe, Ken Gray, Vladimir Gridin, Alex Gurevich, Jorge
Jos\'e, David Larbalestier, Franc Nabarro, Anant Narlikar, Terry
Orlando, Wilson Ortiz, Paulo Pureur, Jacob Schaf, Donglu Shi, James
Thompson and Michael Tinkham are highly appreciated.

\vspace{5mm}

\end{document}